\documentclass{article}
\usepackage{ismir,amsmath,cite,url}
\usepackage[usenames,dvipsnames,svgnames,table]{xcolor}

\usepackage{textcomp} 
\usepackage{booktabs}

\usepackage{subfiles}

\usepackage{rotating}
\usepackage{graphicx}
\usepackage{caption}
\usepackage{subcaption}
\usepackage{makecell}
\usepackage{soul} 
\usepackage{diagbox} 
\usepackage{multirow}

\usepackage{mathtools}
\usepackage{float}
\usepackage{amsmath}

\renewenvironment{description}[1][0pt]
  {\list{}{\labelwidth=.25cm \leftmargin=#1
   }}
  {\endlist}

\graphicspath{{./figs/}}

\newcommand{\MASK}[1]{}

\usepackage{lipsum}

\title{Conditioned-U-Net: introducing a control mechanism in the U-Net for multiple source separations}

\twoauthors
 {Gabriel Meseguer-Brocal} {\small Ircam Lab, CNRS, Sorbonne Universit\'{e} \\ \small  75004 Paris, France  \\
     {\tt \small gabriel.meseguerbrocal@ircam.fr}}
 {Geoffroy Peeters} {\small LTCI, T\'{e}l\'{e}com Paris, Institut Polytechnique de Paris, \\ \small 75013, Paris, France \\
      {\tt \small geoffroy.peeters@telecom-paris.fr}}

\begin{document}

\maketitle

\begin{abstract}

Data-driven models for audio source separation such as U-Net or Wave-U-Net are usually models dedicated to and specifically trained for a single task, e.g. a particular instrument isolation.
Training them for various tasks at once commonly results in worse performances than training them for a single specialized task.
In this work, we introduce the Conditioned-U-Net (C-U-Net) which adds a control mechanism to the standard U-Net.
The control mechanism allows us to train a unique and generic U-Net to perform the separation of various instruments.
The C-U-Net decides the instrument to isolate according to a one-hot-encoding input vector.
The input vector is embedded to obtain the parameters that control Feature-wise Linear Modulation (FiLM) layers.
FiLM layers modify the U-Net feature maps in order to separate the desired instrument via affine transformations.
The C-U-Net performs different instrument separations, all with a single model achieving the same performances as the dedicated ones at a lower cost.

\end{abstract}

\section{Introduction}

Generally, in Music Information Retrieval (MIR) we develop dedicated systems for specific tasks. Facing new (but similar) tasks require the development of new (but similar) specific systems.
This is the case of data-driven music source separation systems.
Source separation aims to isolate the different instruments that appear in an audio mixture (a mixed music track) i.e., reversing the mixing process.
Data-driven methods use supervised learning where the mixture signals and the isolated instruments are available for training.
The usual approach is to build dedicated models for each task to isolate\cite{Jansson_2017, Stoller_2018}.
This has been proved to show great results.
However, since isolating an instrument requires a specific system, we can easily run into problems such as scaling issues (100 instruments = 100 systems).
Besides, these models do not use the commonalities between instruments.
If we modify them to do various tasks at once i.e., adding fix numbers of output masks in last layers, they reduce their performance\cite{Stoller_2018}.

Conditioning learning has appeared as a solution to problems that need the integration of multiple resources of information.
Concretely, when we want to process one in the context of another i.e., modulating a system computation by the presence of external data.
Conditioning learning divides problems into two elements: a \textbf{generic system} and a \textbf{control mechanism} that governs it according to external data.
Although there is a large diversity of domains that use it, it has been developed mainly in the image processing field for tasks such as visual reasoning or style transfer.
There, it has been proved very \-effective, improving the state of the art results\cite{Perez_2017, Vries_2017, Strub_2018}.
This paradigm can be integrated into source separation creating a generic model that adapts to isolate a particular instrument via a control mechanism.
We also believe that this paradigm can benefit to a great diversity of MIR tasks such as multi-pitch estimation, music transcription or music generation.

\begin{figure}[t!]
  \centerline{
    \includegraphics[width=.5\textwidth]{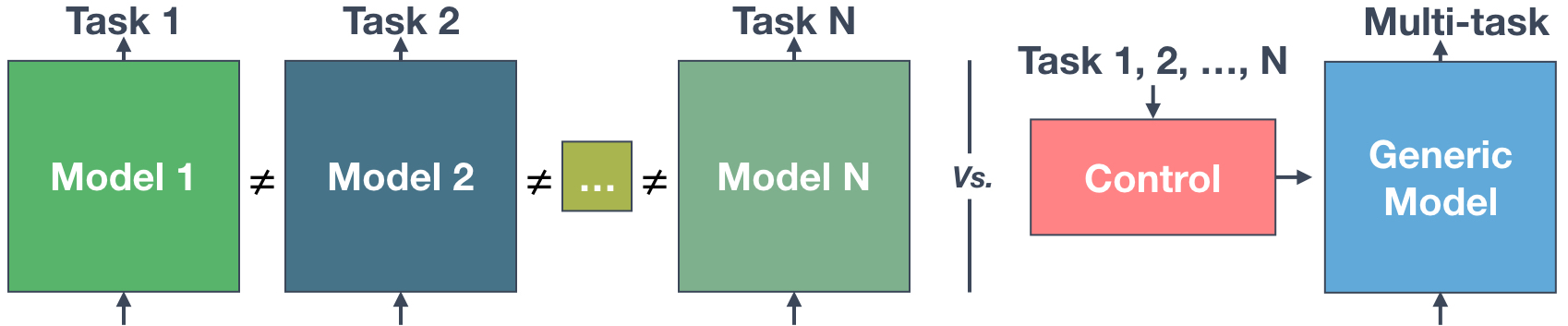}}
  \caption{[Left part] Traditional approach: a dedicated U-Net is trained to separate a specific source.
  [Right part] Our proposition based on conditioning learning. The problem is divided in two: a standard U-Net (which provides generic source separation filters) and a control mechanism.
  This division allows the same model to deal with different tasks using the commonalities between them.
  %
  }
  \label{fig:general-schema}
\end{figure}

In this work, we propose the application of conditioning learning for music source separation.
Our system relies on
a \textbf{standard U-Net system} not specialized in a specific task but rather in finding a set of generic source separation filters, that we
\textbf{control} differently for isolating a particular instrument,
as illustrated in Figure \ref{fig:general-schema}.
Our system takes as input the spectrogram of the mixed audio signal and the control vector.
It gives as output (only one) the separated instrument defined by the control vector.
The main advantages of our approach are
- direct use of commonalities between different instruments,
- a constant number of parameters no matter how many instruments the system is dealing with
- and scalable architecture, in the sense that new instruments can be potentially added without training from scratch a new system.
Our key contributions are:
\begin{enumerate}
  \item the Conditioned-U-Net (C-U-Net), a joint model that changes its behavior depending on external data and performs for any task as good as a dedicated model trained for it. C-U-Net has a fixed number of parameters no matter the number of output sources.
  \item The C-U-Net proves that conditioning learning (via Feature-wise Linear Modulation (FiLM) layers) is an efficient way of inserting external information to MIR problems.
  \item A new FiLM layer that works as good as the original one but with a lower cost (fewer parameters).
\end{enumerate}

\section{Related work}

We review only works related to conditioning in audio and to data-driven source separation methods.

\textbf{Conditioning in audio.}
It has been mainly explored in \textbf{speech generation}.
In the WaveNet approach\cite{Oord_2016, Oord_2017} the speaker identity is fed to a conditional distribution adding a learnable bias to the gated activation units.
A WaveNet modified version is presented in \cite{Shen_2017}. The time-domain waveform generation is conditioned by a sequence of Mel spectrogram computed from an input character sequence (using a recurrent sequence-to-sequence network with attention).
In \textbf{speech recognition} conditions are used in \cite{Kim_2017}, applying conditional normalisation to a deep bidirectional LSTM (Long Short Term Memory) for dynamically generating the parameters in the normalisation layer.
This model adapts itself to different acoustic scenarios.
In \cite{Kim_2017}, the conditions do not come from any external source but rather from utterance information of the model itself.
They have been also used in \textbf{music generation} for accompaniments conditioned on melodies\cite{Cheng-Zhi_2018} or incorporating history information (melody and chords) from previous measures in a generative adversarial network (GAN)\cite{Yang_2017}.
Finally, it has been also proved to be very efficient for \textbf{piano transcription} \cite{Hawthorne_2018}: the pitch onset detection is internally concatenated to the frame-wise pitch prediction controlling if a new pitch starts or not. Both, onset detection and frame-wise prediction are trained together.

\textbf{Source separation based on supervised learning.}
We refer the reader to \cite{Rafii_2018} for an extensive overview of the different source separation techniques.
We review only the data-driven approaches.
Here, the neural networks have taken the lead.
Although architectures such as RNN\cite{Huang_2015} or CNN\cite{Chandna_2017} have been studied, the most successful one use a deep U-Net architecture (also called U-Net).
In \cite{Jansson_2017}, the U-Net is applied to a spectrogram to separate the vocal and accompaniment components, training a specific model for each task.
Since the output is the spectrogram, they need to reconstruct the audio signal which potentially leads to artifacts.
For this reason, Wave-U-Net proposes to apply the U-Net to the audio-waveform\cite{Stoller_2018}.
They also adapt their model for isolating different sources at once by adding to their dedicated version as many outputs as sources to separate.
However, this multi-instruments version performs worse than the dedicated one (for vocal isolation) and has to be retrained to different source combinations.

The closest work to ours is \cite{Kameoka_2018}.
In there, they propose to use multi-channel audio as input to a Variational Auto-Encoder (VAE) to separate 4 different speakers.
The VAE is conditioned on the ID of the speaker to be separated.
The proposed method outperforms its baseline.

\section{Conditioning learning methodology}
\label{sec:methodology}

\begin{figure}[t!]
  \centerline{
    \includegraphics[width=.4\textwidth]{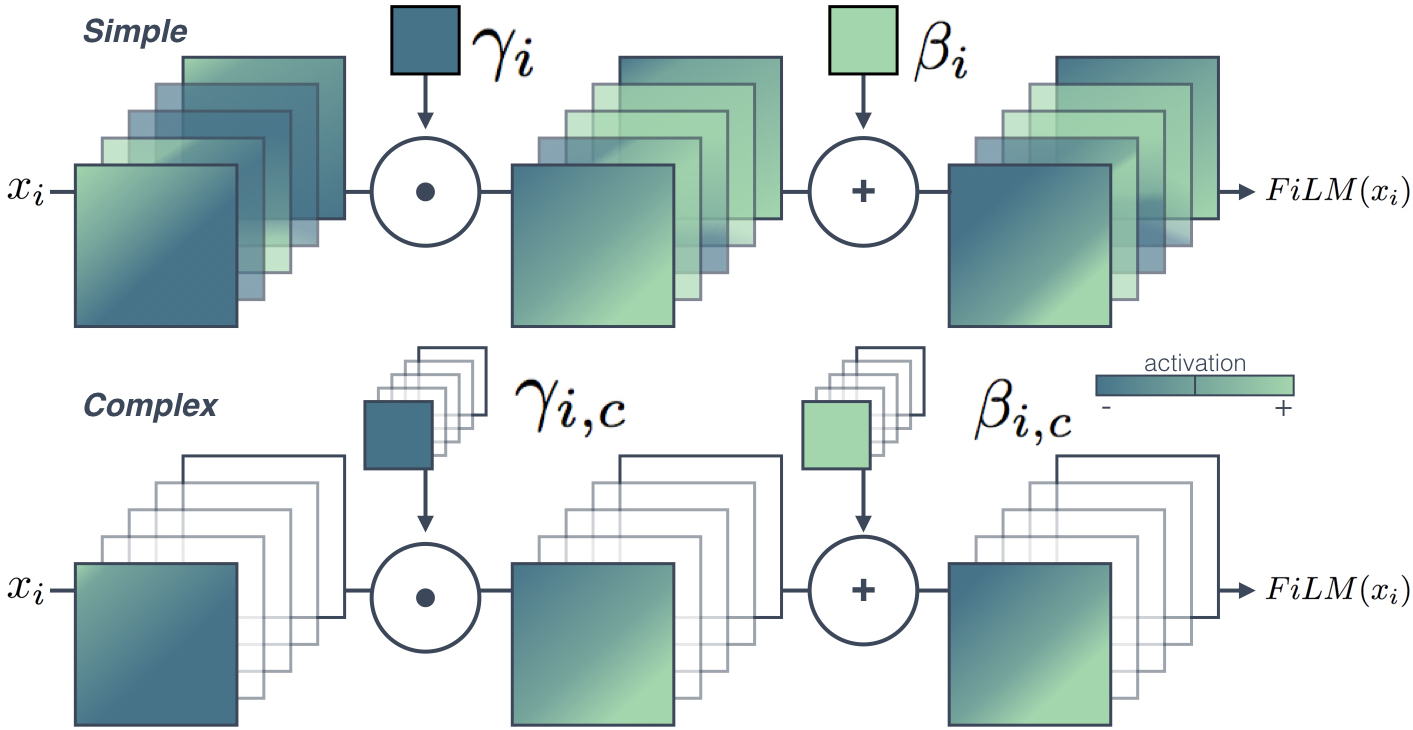}}
  \caption{[Top part] \textit{FiLM simple} layer applies the same affine transformation to all the input feature maps $x$.
  [Bottom part] In the \textit{FiLM complex} layer, independent affine transformations are applied to each feature map $c$.}
  \label{fig:film}
\end{figure}

\subsection{Conditioning mechanism.}
There are many ways to condition a network (see \cite{Dumoulin_2018} for a wide overview) but most of them can be formalized as affine transformations denoted by the acronym FiLM (Feature-wise Linear Modulation) \cite{Perez_2017}.
FiLM permits to modulate any neural network architecture inserting one or several FiLM layers at any depth of the original model.
A FiLM layer conditions the network computation by applying an affine transformation to intermediate features:

\begin{equation}
    FiLM(x) = \gamma(z) \cdot x + \beta(z)
\end{equation}

\noindent where $x$ is the input of the FiLM layer (i.e., the intermediate feature we want to modify), $\gamma$ and $\beta$ are parameters to be learned.
They scale and shift $x$ based on the external information, $z$.
The output of a FiLM layer has the same dimension as the intermediate feature input $x$.
FiLM layers can be inserted at any depth $i$ in the controlled network.

As described in Figure \ref{fig:film},
the original FiLM layer applies an independent affine transformation to each feature map $c$\footnote{Or element-wise.}:  $\gamma_{i,c}$ and $\beta_{i,c}$ \cite{Perez_2017}.
We call this a \textit{FiLM complex} layer (\textbf{Co}).
We propose a simpler version that applies the same $\gamma_{i}$ and $\beta_{i}$ to all the feature maps (therefore $\gamma$ and $\beta$ do not depend on $c$).
We call it a \textit{FiLM simple} layer (\textbf{Si}).
The \textit{FiLM simple} layer decreases the degrees of freedom of the transformations to be carried out forcing them to be generic and less specialized. It also reduces drastically the number of parameters to be trained.
As FiLM layers do not change the shape of $x$, FiLM is transparent and can be used in any particular architecture providing flexibility to the network by adding a control mechanism.

\subsection{Conditioning architecture.}
A conditioning architecture has two components:
\begin{description}
  \item[The conditioned network.]
  It is the network that carries out the core computation and obtains the final output.
  It is usually a generic network that we want to behave differently according to external data.
  Its behavior is altered by the condition parameters, $\gamma_{i,(c)}$ and $\beta_{i,(c)}$ via FiLM layers.

  \item[The control mechanism - condition generator.]
  It is the system that produces the parameters ($\gamma$'s and $\beta$'s) for the FiLM layers with respect to the external information $z$: the input conditions.
  It codifies the task at hand and provides the instructions to control the conditioned network.
  The condition generator can be trained jointly\cite{Perez_2017, Strub_2018} or separately with the conditioned network\cite{Vries_2017}.
\end{description}

This paradigm clearly separates the tasks description and control instructions from the main core computation.

\section{Conditioned-U-Net for multitask source separation}

\newcommand{\CMCG}[0]{control mechanism/condition generator }

We formalize source separation as a multi-tasks problem where one task corresponds to the isolation of one instrument.
We assume that while the tasks are different they share many similarities, hence they will benefit from a conditioned architecture.
We name our approach the \textbf{Conditioned-U-Net} (C-U-net).
It differs from the previous works where a dedicated model is trained for a single task\cite{Jansson_2017} or where it has a fixed number of outputs\cite{Stoller_2018}.

\begin{figure}[t!]
  \centerline{
    \includegraphics[width=.4\textwidth]{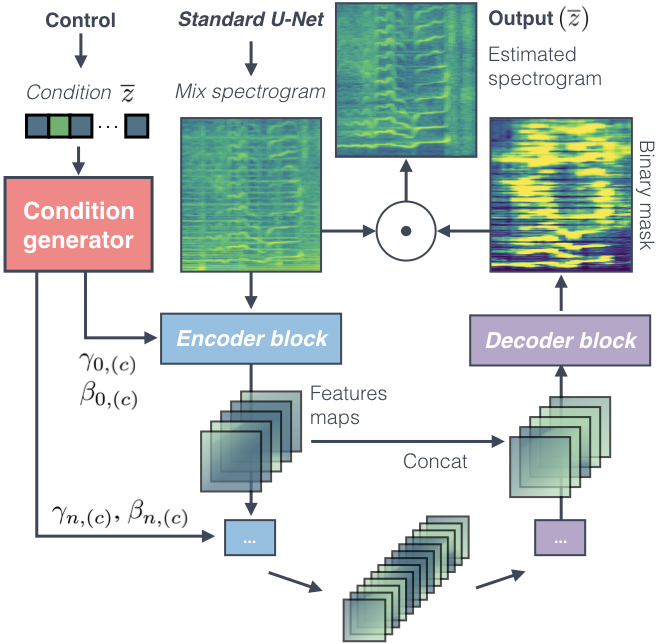}}
    \caption{The C-U-Net has two distinct parts: the condition generator and a standard U-Net. The former codifies the input the condition vector, $\overline{z}$ (with the instrument to isolate) for getting the needed $\gamma_{i,(c)}$ and $\beta_{i,(c)}$. The generic U-Net has as input the magnitude spectrum. It adapts its conduct via FiLM layers inserted in the encoder part. The system outputs the desired instrument defined by $\overline{z}$.}
  \label{fig:C-U-Net}
\end{figure}

As in \cite{Jansson_2017, Stoller_2018}, our \textbf{conditioned network} is a standard U-Net that computes a set of generic source separation filters that we use to separate the various instruments.
It adapts itself through the control mechanism (the condition generator) with FiLM layers inserted at different depths.
Our external data is a condition vector $\overline{z}$ (a one-hot-encoding) which specify the instrument to be separated.
For example, $\overline{z}=[0,1,0,0]$ corresponds to the drums.
The vector $\overline{z}$ is the input to the \CMCG that has to learn the best  $\gamma_{i,c}$ and $\beta_{i,c}$ values such that, when they modify the feature maps (in the FiLM layers) the C-U-Net separates the indicated instrument i.e., it decides which features maps information is useful to get each instrument.
The \CMCG is itself a neural network that embeds $\overline{z}$ into the best $\gamma_{i,c}$ and $\beta_{i,c}$.
The conditioned network and the condition generator are trained jointly.
A diagram is shown in Figure\ref{fig:C-U-Net}.

Our C-U-Net can perform different instrument source separations as it alters its behavior depending on the value of the external condition vector $\overline{z}$.
The inputs of our system are the mixture and the vector $\overline{z}$.
There is only one output, which corresponds to the isolated instrument defined by $\overline{z}$.
While training, the output corresponds to the desired isolated instrument that matches the $\overline{z}$ activation.

\subsection{Conditioned network: U-Net architecture}

We used the U-Net architecture proposed for vocal separation\cite{Jansson_2017}, which is an adaptation of the microscopic images U-Net\cite{Ronneberger_2015}. 
The input and output are magnitude spectrograms of the monophonic mixture and the instrument to isolate.
The U-Net follows an encoder-decoder architecture and adds a skip connection to it.

\begin{description}
  \item[Encoder.] It creates a compressed and deep representation of the input by reducing its dimensionality while preserving the relevant information for the separation.
  It consists of a stack of convolutional layers, where each layer halves the size of the input but doubles the number of channels.
  \item[Decoder.] It reconstructs and interprets the deep features and transforms it into the final spectrogram.
  It consists of a stack of deconvolutional layers.
  \item[Skip-connections.]
  As the encoder and decoder are symmetric i.e., feature maps at the same depth have the same shape, the U-Net adds skip-connections between layers of the encoder and decoder of the same depth.
  This refines the reconstruction by progressively providing finer-grained information from the encoder to the decoder.
  Namely, feature maps of a layer in the encoder are concatenated to the equivalent ones in the decoder.
\end{description}

The final layer is \textbf{a soft mask} (sigmoid function $\in [0,1]$) $f(X, \theta)$ which is applied to the input $X$ to get the isolated source $Y$. The loss of the U-Net is defined as:
  \begin{equation}
    \mathcal{L}(X, Y; \theta) = \|f (X, \theta) \odot X - Y \|_{1,1}
  \end{equation}
 where $\theta$ are the parameters of the system.

\textbf{Architecture details.}
Our implementation mimics the original one\cite{Jansson_2017}.
The \textbf{encoder} consists in 6 encoder blocks.
Each one is made of a 2D convolution with 5x5 filters, stride 2, batch normalisation, and leaky rectified linear units (ReLU) with leakiness 0.2.
The first layer has 16 filters and we double them for each new block.
The \textbf{decoder} maps the encoder, with 6 decoders blocks with stride deconvolution, stride 2 and a 5x5 kernel, batch normalisation, plain ReLU, and a 50\% dropout in the first three.
The final one, the soft mask, uses a sigmoid activation.
The model is trained using the ADAM optimiser \cite{Diederik_2014} and a 0.001 learning rate.
As in \cite{Jansson_2017}, we downsample to 8192 Hz, compute the Short Time Fourier Transform with a window size of 1024 and hop length of 768 frames.
The input is a patch of 128 frames (roughly 11 seconds) from the normalised (per song to [0, 1]) magnitude spectrogram for both the mixture spectrogram and the isolated instrument.

\begin{figure}[t!]
  \centerline{
    \includegraphics[width=.4\textwidth]{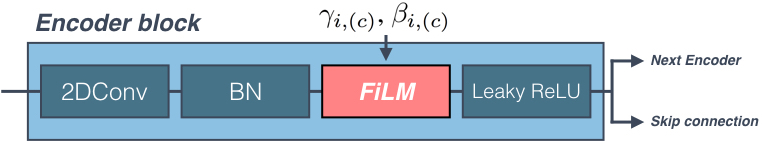}}
    \caption{FiLM layers are placed after the batch normalisation. The output of a encoding block is connected to both, the next encoding block and the equivalent layer in the decoder via the skip connections.}
  \label{fig:encoding-block}
\end{figure}

\textbf{Inserting FiLM.} The U-Net has two well differentiated stages: the \textbf{encoder} and \textbf{decoder}.
The \textbf{enconder} is the part that transforms the mixture magnitude input into a deep representation capturing the key elements to isolate an instrument.
The \textbf{decoder} interprets this representation for reconstructing the final audio.
We hypothesise that, if we can have a different way of encoding each instrument i.e., obtaining different deep representations, we can use a common `universal' decoder to interpret all of them.
Following this reasoning, we decided to condition only the U-Net encoder part.
In the C-U-Net, a FiLM layer is inserted inside each encoding block after the batch normalisation and before the Leaky ReLU, as described in Figure \ref{fig:encoding-block}.
This decision relies on previous works where feature are modified after the normalisation\cite{Perez_2017, Vries_2017, Kim_2017}.
Batch normalisation normalises each feature map so that it has zero mean and unit variance \cite{Ioffe_2015}.  Applying FiLM after batch normalisation re-scale and re-shift feature maps after the activations.
This allows the net to specialise itself to different tasks.
As the output of our encoding blocks is transformed by the FiLM layer the data that flows through the skip connections carries on also the transformations.
If we use the \textit{FiLM complex} layer, the \CMCG needs to generate 2016 parameters (1008 $\gamma_{i,c}$ and 1008 $\beta_{i,c}$). On the other hand, \textit{FiLM simple} layers imply 12 parameters: one $\gamma_{i}$ and one $\beta_{i}$ for each of the 6 different encoding blocks, which means 2002 parameters less than for \textit{FiLM complex} layers.

\subsection{Condition generator: Embedding nets}

The \CMCG computes the $\gamma_{i,(c)}(\overline{z})$ and $\beta_{i,(c)}(\overline{z})$ that modify our standard U-net behavior.
Its architecture has to be flexible and robust to generate the best possible parameters.
It has also to be able to find relationships between instruments.
That is to say, we want it to produce similar $\gamma_{i,(c)}$ and $\beta_{i,(c)}$ for instruments that have similar spectrogram characteristics.
Hence, we explore two different embeddings: a fully connected version and a convolutional one (CNN).
Each one is adapted for the \textit{FiLM complex} layer as well as for the \textit{FiLM simple} layer.
In every \CMCG configuration, the last layer is always two concatenated fully connected layers. Each one has as many parameters ($\gamma$'s or $\beta$'s) as needed. With this distinction we can control $\gamma_{i,(c)}$ and $\beta_{i,(c)}$ individually (different activations).

\begin{description}
  \item [\textbf{Fully-Connected embedding (F)}:] it is formed of a first dense layer of 16 neurons and two fully connected blocks (dense layer, 50\% dropout and batch normalization) with 64 and 256 neurons for \textit{FiLM simple} and 256 and 1024 for \textit{FiLM complex}. All the neurons have relu activations. The last fully connected block is connected with the final \CMCG layer i.e., the two fully connected ones ($\gamma_{i,c}$ and $\beta_{i,c}$).
  We call the C-U-Net that uses these architectures \textit{C-U-Net-SiF} and \textit{C-U-Net-CoF}.
  \item [\textbf{CNN embedding (C)}:] similarly to the previous one and inspired by \cite{Shen_2017}, this embedding consists in a 1D convolution with $lenght(\overline{z})$ filters followed by two convolution blocks (1D convolution with also $lenght(\overline{z})$ filters, 50\% dropout and batch normalization). The first two convolutions have `same' padding and the last one, `valid'. Activations are also relu. The number of filters are 16, 32 and 64 for the \textit{FiLM simple} version and 32, 64, 256 for the \textit{FiLM complex} one.
  Again, the last CNN block is connected with the two fully connected ones.
  The C-U-Net that uses these architectures are called \textit{C-U-Net-SiC} and \textit{C-U-Net-CoC}.
  This embedding is specially designed for dealing with several instruments because it seems more appropriated to find common $\gamma_{i,(c)}$ and $\beta_{i,(c)}$ values for similar instruments.
\end{description}

\begin{table}[t]
  \centering
  \caption{Params number in millions. With dedicated U-Nets, each task needs a model with 10M params. C-U-Nets are multi-task and the number of params remains constant.}
  \label{table:cunet-overview}
  \resizebox{.5\textwidth}{!}{%
  \small
  \begin{tabular}{| c | c | c | c | c | c | }
    \hline
        MODEL      &  Non-conditioned      &  SiF     & CoF    & SiC    & CoC    \\
    \hline
        PARAM      &   39,30   (4 tasks x 9,825)   &   9,85     & 12   & 9,84    & 10,42         \\

    \hline
  \end{tabular}%
  }
\end{table}

The various control mechanisms only introduce a reduced number of parameters to the standard U-Net architecture remaining constant regardless of the instruments to separate, Table \ref{table:cunet-overview}.
Additionally, they make direct use of the commonalities between instruments.

\section{Evaluation}

Our objective is to prove that conditioned learning via FiLM (generic model+control) allows us to transform the U-Net into a multi-task system without losing performances.
In Section \ref{sec:miscellaneous} we review our experiment design aspects and we detail the experiment to validate the multi-task capability of the C-U-Net in Section \ref{sec:exp-multitask}.

\subsection{Evaluation protocol}
\label{sec:miscellaneous}

\begin{table}[t!]
  \centering
  \caption{Overall performance (mean $\pm$ std) for the 4 tasks.
   \textit{Si}= simple FiLM, \textit{Co}= complex FiLM,  \textit{F}= Fully-embed and \textit{C}= CNN-embed,  \textit{p}= progressive train or \textit{np}= not.}
  \label{table:cunet-multitask}
  \resizebox{.5\textwidth}{!}{%
  \small
  \begin{tabular}{| c  | c|c|c |}
    \hline
     \multirow{2}{*}{MODEL}  & \multicolumn{3}{c|}{Total} \\ \cline{2-4}
      &  SIR   &  SAR  &  SDR     \\
    \hline

    \textit{Fix-U-Net(x4)}             & 7.31 $\pm$ 4.04 & 5.70 $\pm$ 3.10 & 2.36 $\pm$ 3.96 \\
    \hline
    \hline
    \textit{C-U-Net-SiC-np}            & 7.35 $\pm$ 4.13 & 5.74 $\pm$ 3.18 & 2.34 $\pm$ 3.69 \\
    \textit{C-U-Net-SiC-p}             & \textbf{8.00} $\pm$ 4.37 & \textbf{5.74} $\pm$ 3.63 & \textbf{2.54} $\pm$ 4.07  \\
    \hline
    \textit{C-U-Net-CoC-np}            & 7.27 $\pm$ 4.24 & 5.60 $\pm$ 2.88 & 2.36 $\pm$ 3.81  \\
    \textit{C-U-Net-CoC-p}             & 7.49 $\pm$ 4.54 & 5.67 $\pm$ 3.03 & 2.42 $\pm$ 4.21  \\
    \hline
    \textit{C-U-Net-SiF-np}            & 7.23 $\pm$ 3.97 & 5.59 $\pm$ 3.01 & 2.22 $\pm$ 3.67 \\
    \textit{C-U-Net-SiF-p}             & 7.64 $\pm$ 4.05 & 5.73 $\pm$ 2.88 & 2.46 $\pm$ 3.88 \\
    \hline
    \textit{C-U-Net-CoF-np}            & 7.42 $\pm$ 4.20 & 5.59 $\pm$ 3.07 & 2.32 $\pm$ 3.85 \\
    \textit{C-U-Net-CoF-p}             & 7.52 $\pm$ 4.04 & 5.71 $\pm$ 2.99 & 2.42 $\pm$ 3.97 \\

    \hline
  \end{tabular}%
  }
\end{table}

\begin{description}
  \item[Dataset.]
  We use the Musdb18 dataset \cite{musdb_2018}.
  It consists of 150 tracks with a defined split of 100 tracks for training and 50 for testing.
  From the 100 tracks, we use 95 (randomly assigned) for training, and the remaining 5 for the validation set, which is used for early stopping.
  The performance is evaluated on the 50 test tracks.
  In Musdb18, mixtures are divided into four different sources: \textbf{Vocals}, \textbf{Bass}, \textbf{Drums} and \textbf{Rest} of instruments. The 'Rest' task mixes every instrument that it is not vocal, bass or drums.
  Consequently, the C-U-Net is trained for four tasks (one task per instrument) and $\overline{z}$ has four elements.

  \item[Evaluation metrics.]
  We evaluate the performances of the separation using the mir evaltoolbox \cite{Raffel_2014}.
  We compute three metrics: Source-to-Interference Ratios (SIR), Source-to-Artifact Ratios (SAR) and Source-to-Distortion Ratios (SDR)\cite{Vincent_2006}.
  To compute the three measure we also need the predicted 'accompaniment' (the mixture part that does not correspond to the target source). Each task has a different accompaniment e.g., for the drums the accompaniment is rest+vocals+bass. We create the accompaniments by adding the audio signal of the needed sources.

  \item[Audio Reconstruction method.]
  The system works exclusively on the magnitude of audio spectrograms.
  The output magnitude is obtained by applying the mask to the mixture magnitude.
  As in \cite{Jansson_2017}, the final predicted source (the isolated audio signal) is reconstructed concatenating temporally (without overlap) the output magnitude spectrums and using the original mix phase unaltered.
  We compute the predicted accompaniment subtracting the predicted isolated signal to the original mixture.
  Despite there are better phase reconstruction techniques such as \cite{Mayer_2017}, errors due to this step are common to both methods (U-Net and C-U-Net) and do not affect our main goal: to validate conditioning learning for source separation.

  \item[Activation function for $\gamma$ and $\beta$.]
  One of the most important design choices is the activation function for $\gamma_{i,(c)}$ and $\beta_{i,(c)}$.
  We tested all the possible combinations of three activation \-functions (linear, sigmoid and tanh) in the \textit{C-U-Net-SiF} configuration.
  As in \cite{Perez_2017}, the C-U-Net works better when $\gamma_{i,(c)}$ and $\beta_{i,(c)}$ are linear.
  Hence, our $\gamma$'s and $\beta$'s have always linear \-activations.

  \item[Training flexibility.] The conditioning mechanism gives the flexibility to have continuous values in the input $\overline{z} \in [0,1]$, which weights the target output $Y$ by the same value.
  We call this training method \textbf{progressive}.
  In practice, while training, we randomly weight $\overline{z}$ and $Y$ by a value between 0 and 1 every 5 instances.
  This is a way of dealing with ablations by making the control mechanism robust to noise.
  As shown in Table \ref{table:cunet-multitask}, this training procedure \textit{(p)} improves the models.
  Thus, we adopt it in our training.
  Moreover, preliminary results (not reported) show that the C-U-Net can be trained for complex tasks like bass+drums or voice+drums.
  These complex tasks could benefit from `in between-class learning' method\cite{Tokozume_2017} where $\overline{z}$ will have different intermediate instrument combinations.

\end{description}

\begin{table}[t!]
  \centering
  \caption{Task comparison between the \textit{dedicated U-Nets} and the \textit{C-U-Net-CoF}. Results indicate that they perform similarly for all the tasks. We also add the multi-instrument Wave-U-Net (M) results (median in parenthesis) and when possible the dedicated version (D). For vocals isolation the Wave-U-Net-M performs worse than the Wave-U-Net-D.}
  \label{table:cunet-multitask-details}
  \resizebox{.5\textwidth}{!}{%
  \small
  \begin{tabular}{| c | c  || c|c|c |}
    \hline
     & Model                            &   SIR   &  SAR  &  SDR     \\
    \hline

    \hline
    \parbox[t]{2mm}{\multirow{4}{*}{\rotatebox[origin=c]{90}{Vocals}}}
    & \textit{Fix-U-Net(x4)}             & 10.70 $\pm$ 4.26  & 5.39 $\pm$ 3.58 & 3.52 $\pm$ 4.88  (4.72) \\
    & \textit{C-U-Net-CoF}               & 10.76 $\pm$ 4.39  & 5.32 $\pm$ 3.27 & 3.50 $\pm$ 4.37  (4.65) \\
    & \textit{Wave-U-Net-D}              & -                 &  -              & 0.55 $\pm$ 13.67 (4.58) \\
    & \textit{Wave-U-Net-M}              & -                 &  -              & -2.10 $\pm$ 15.41 (3.0) \\
    \hline
    \parbox[t]{2mm}{\multirow{3}{*}{\rotatebox[origin=c]{90}{Drums}}}
    & \textit{Fix-U-Net(x4)}             & 10.08 $\pm$ 4.28  & 6.42 $\pm$ 3.28 & 4.28 $\pm$ 3.65 (4.13) \\
    & \textit{C-U-Net-CoF}               & 10.03 $\pm$ 4.34  & 6.80 $\pm$ 3.25 & 4.30 $\pm$ 3.81 (4.38) \\
    & \textit{Wave-U-Net-M}              & -                 &  -              & 2.88 $\pm$ 7.68 (4.15) \\
    \hline
    \parbox[t]{2mm}{\multirow{3}{*}{\rotatebox[origin=c]{90}{Bass}}}
    & \textit{Fix-U-Net(x4)}             & 4.64 $\pm$ 4.76   & 6.51 $\pm$ 2.68 & 1.46 $\pm$ 4.31 (2.48)  \\
    & \textit{C-U-Net-CoF}               & 5.30 $\pm$ 4.73   & 6.29 $\pm$ 2.39 & 1.65 $\pm$ 4.07 (2.60)  \\
    & \textit{Wave-U-Net-M}              & -                 &  -              & -0.30 $\pm$ 13.50 (2.91) \\
    \hline
    \parbox[t]{2mm}{\multirow{3}{*}{\rotatebox[origin=c]{90}{Rest}}}
    & \textit{Fix-U-Net(x4) }            & 3.83 $\pm$ 2.84   & 4.47 $\pm$ 2.85 & 0.19 $\pm$ 3.00 (0.97) \\
    & \textit{C-U-Net-CoF}               & 4.00 $\pm$ 2.70   & 4.37 $\pm$ 3.06 & 0.24 $\pm$ 3.64 (1.71) \\
    & \textit{Wave-U-Net-M}              & -                 &  -              & 1.68 $\pm$ 6.14 (2.03) \\
    \hline
  \end{tabular}%
  }
\end{table}

\subsection{Multitask experiment}
\label{sec:exp-multitask}

\begin{figure*}[t!]
  \centerline{
    \includegraphics[width=.95\textwidth]{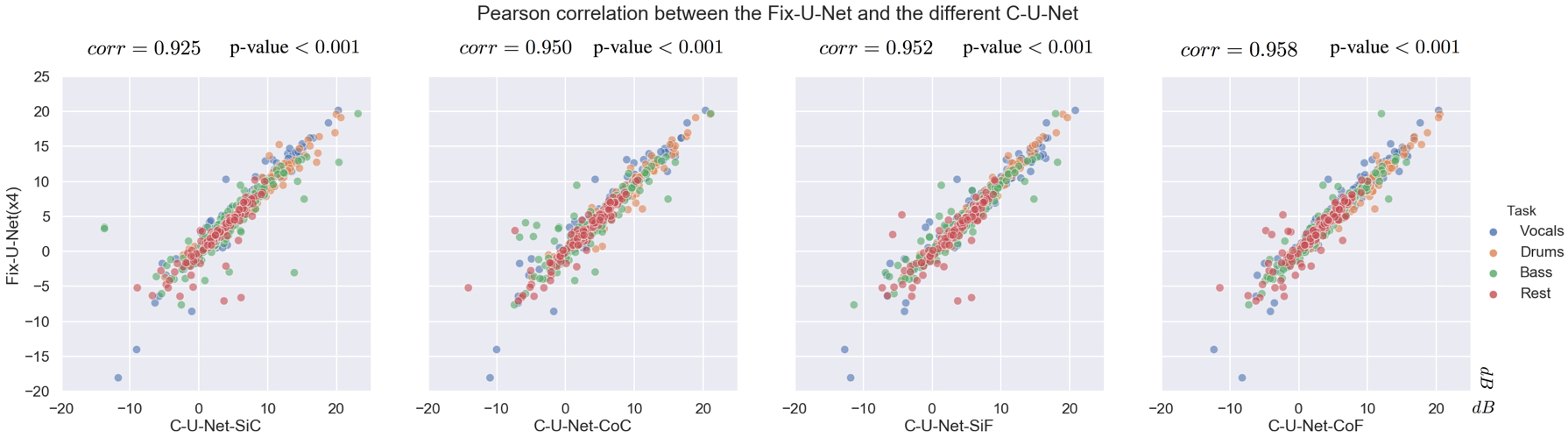}}
  \caption{
  Each graph correlates the performance of two models. On top of it, we show the correlation and p-value. The 'y' axis represents the fixed version (the four dedicated U-Nets) and the 'x' one a different C-U-Net version (with progressive train).
  The coordinates of each dots correspond to the models' performance i.e., 'y' position for the Fix-U-Net performance and 'x' for C-U-Net.
  There are three dots per song one per metric (SIR, SAR, and SDR) which does a total of 600 (50 songs x 3 metrics x 4 instruments).
  The dots alignment in the diagonal implies a strong correlation between models: if one works well, the others too and vice versa.
  Each color highlights the points of each source separation task.}
  \label{fig:corr}
\end{figure*}

We want to prove that a given C-U-Net can isolate the \textbf{Vocals}, \textbf{Drums}, \textbf{Bass}, and \textbf{Rest} as good as four dedicated U-Net trained specifically for each task\footnote{with the same learning rate and optimizer as the C-U-Nets.}
We call this set of dedicated U-Nets, \textit{Fix-U-Nets}.
Each C-U-Nets version (one model) is compared with the Fix-U-Nets set (four models).
We review the results at Table \ref{table:cunet-multitask} and show a comparison per task in Table \ref{table:cunet-multitask-details}.

Results in Table \ref{table:cunet-multitask} for all 4 instruments highlight that \textit{FiLM simple} layers work as good as the complex ones.
This is quite interesting because it means that applying 6 affine transformations with just 12 scalars (6 $\gamma_{i}$ and 6 $\beta_{i}$) at a precise point allows the C-U-Net to do several source separations.
With \textit{FiLM complex} layers it is intuitive to think that treating each feature map individually let the C-U-Net learn several deep representations in the encoder.
However, we have no intuitive explanation for \textit{FiLM simple} layers.
We did the Tukey test with no significant differences between the \textit{Fix-U-Nets} and the \textit{C-U-Nets} for any task and metric.
Another remark is that the four C-U-Nets benefit from the \textit{progressive} training.
Nevertheless, it impacts more the simple layers than in the complex ones.
We think that the restriction of the former (fewer parameters) helps them to find an optimal state.

However, these results do not prove nor discard the significant similarity between systems.
For demonstrating that we have carried out a Pearson correlation experiment.
The results are detailed in Figure \ref{fig:corr}.
The Pearson coefficient measures the linear relationship between two sets of results (+1 implies an exact linear relationship).
It also computes the p-value that indicates the probability that uncorrelated systems have produces them.
Our distinct C-U-Net configurations have a global $corr > .9$ and $\textrm{p-value} < 0.001$.
Which means that there is always more than $90\%$ correlation between the performance of the four dedicated \textit{U-Nets} and the (various) conditional version(s).
Additionally, there is almost no probability that a C-U-Net version is not correlated with the dedicated ones.
We have also computed the Pearson coefficient and p-value per task and per metric with the same results.
In Figure \ref{fig:corr} shows a strong correlation between the \textit{Fix-U-Net} results and the distinct \textit{C-U-Nets} (independently of the task or metric). Thus, if one works well, the others too and vice versa.

In Table \ref{table:cunet-multitask-details} we detail the results per task and metric for the \textit{Fix-U-Net} and the \textit{C-U-Net-CoF} which is not the best C-U-Net but the one with the highest correlation with the dedicated ones.
There we can see how their performances are almost identical.
Nevertheless, our vocal isolation (in any case) is not as good as the one reported in \cite{Jansson_2017}, we believe that this is mainly due to the lack of data.
These results can only be compared with the Wave-U-Net\cite{Stoller_2018}.
Although they report the results (only the SDR) for the four tasks in the multi-instrument version (multiple outputs layers) they only have a dedicated version for vocals.
For vocal separation, the performance of the multi-instrument version decreases more than 2.5 dB in mean, 1.5 dB in the median and the std increase in almost 2 dB.
Furthermore, the C-U-Net performs better than the multi-instrument in three out of four tasks (vocals, bass, and drums)\footnote{Our experiment conditions are different in training data size (95 Vs 75) and in sampling rate (8192 Hz Vs 22050 Hz) than Wave-U-Net.}.
For the 'Rest' task, the multi-instrument wave-u-net outperforms our C-U-Nets.
This is normal because the dedicated U-Net has already problems with this class and the C-U-Nets inherits the same issues.
We believe that they come from the vague definition of this class with many different instruments combinations at once.

This proves that the various C-U-Nets behave in the same way as the dedicated U-Nets for each task and metric.
It also demonstrates that conditioned learning via FiLM is robust to diverse control mechanisms/condition generators and FiLM layers.
Moreover, it does not introduce any limitations which are due to other factors.

\section{Conclusions and future work}
We have applied conditioning learning to the problem of instrument source separations by adding a control mechanism to the U-Net architecture. The C-U-Nets can do several source separation tasks without losing performance as it does not introduce any limitation and makes use of the commonalities of the distinct instruments. It has a fixed number of parameters (much lower than the dedicated approach) independently of the number of instruments to separate.
Finally, we showed that \textit{progressive training} improves the C-U-Nets and introduced the \textit{FiLM simple}, a new conditioning layer that works as good as the original one but requires less $\gamma$'s and $\beta$'s.

Conditioning learning faces problems providing a generic model and a control mechanism.
This gives flexibility to the systems but introduces new challenges.
We plan to extend the C-U-Net to more instruments to find its limitations and to explore the performance for complex tasks i.e., separating several instruments combinations (e.g., vocals+drums).
Likewise, we are exploring ways of adding new conditions (namely new instrument isolation) to a trained C-U-Net and how to detach the joint training.
We also intend to integrate it other architectures such as Wave-U-Net and data augmentation techniques \cite{Cohen_2019}.


Lastly, we believe that conditioning learning via FiLM will benefit many MIR problems because it defines a transparent and direct way of inserting external data to modify the behavior of a network.


\pagebreak

\textbf{Acknowledgement.}
This research has received funding from the French National Research Agency under the contract ANR-16-CE23-0017-01 (WASABI project).
Implementation and audio examples available at https://github.com/gabolsgabs/cunet

\small
\bibliography{ismir2019}
\end{document}